\pdfoutput=1
\documentclass[11pt]{article}
\usepackage{graphicx,color} 
\usepackage{jheppub}
\usepackage{amsmath}
\usepackage{amssymb}
\usepackage{relsize}
\usepackage{xcolor}
\usepackage{enumerate}
\usepackage{mathtools}
\usepackage{comment}
\usepackage[utf8]{inputenc}
\usepackage{tikz}
\usepackage{parskip}
\DeclareMathOperator\Disc{Disc}

\newcommand{\beq}{\begin{equation}}
\newcommand{\eeq}{\end{equation}}
\def\d{\delta}
\def\st{\tilde{s}}
\def\dd{\mathrm{d}}
\def\xp{x^{\perp}}
\def\xv{\Vec{x}}
\def\D{\Delta}

\def\f{\phi}
\def\F{\Phi}

\def\g{\gamma}
\def\G{\Gamma}
\def\a{\alpha}

\newcommand{\nn}{\nonumber\\}

\title{Boundary conformal field theory, holography and bulk locality}

\author[a]{Pinaki Banerjee} \emailAdd{pinaki.banerjee@iitgn.ac.in}
\author[b]{\!\!, Parijat Dey} \emailAdd{parijat.dey@bose.res.in}
\author[b]{and Dipyendu Dhar} \emailAdd{dipyendu.dhar@bose.res.in}

\affiliation[a]{Indian Institute of Technology Gandhinagar, Department of Physics, Gujarat 382355, India}	
\affiliation[b]{Department of Astrophysics and High Energy Physics,\\
S.N. Bose National Centre for Basic Sciences,
Salt Lake, Kolkata 700106, India}

\begin{document}

%%%%%%%%%%%%%%%%%%%%%%%%%%%%%%%%%%%%
\abstract{We study bulk locality in a scalar effective field theory (EFT) in AdS background in presence of an end-of-the-world (EOW) brane. The holographic dual description is given in terms of a boundary conformal field theory (BCFT).  
We compute the two point correlation function of scalar operators in the BCFT using the one-loop Witten diagrams and compare its analytic structure with the constraints imposed by boundary conformal symmetry. We find that the loop-corrected correlator derived from a local bulk description is not fully compatible with BCFT expectations. This result places nontrivial constraint on bulk locality in holographic BCFT constructions and identifies BCFT correlators as sensitive probes of quantum bulk dynamics in presence of boundaries. }

\maketitle

%%%%%%%%%%%%%%%%%%%%%%%%%%%%%%%%%%%%
%%%%%%%%%%%%%%%%%%%%%%%%%%%%%%%%%%%%
\section{Introduction}

One of the central lessons of the AdS/CFT correspondence \cite{Maldacena:1997re} is that a higher dimensional space-time with local dynamics can emerge from a lower dimensional conformal field theory (CFT). In the large-$N$, strongly coupled regime, correlation functions in a $d$ dimensional local CFT reorganize in a manner consistent with a bulk description governed by an effective field theory (EFT) on AdS$_{d+1}$. At long wavelengths, this EFT is local: interactions are described by local vertices, propagators have standard short-distance singularities, and correlators exhibit the analytic structure expected of a local quantum field theory in curved spacetime. This emergence of locality is highly nontrivial, given that the CFT itself is defined without reference to space-time locality in the bulk.

A crucial conceptual distinction in this context is between \emph{coarse locality} and \emph{sharp locality}. Coarse locality refers to the statement that bulk physics is local on scales comparable to the AdS radius and is visible already at tree level in the bulk through lower point correlators. Sharp locality is a much stronger requirement: it asserts that the bulk admits a local EFT description at parametrically short distances and that this description remains valid order by order in the $1/N$ expansion. Sharp locality is encoded not merely in the existence of a bulk Lagrangian, but in the detailed analytic structure of CFT correlators -- specifically, in particular kinematic limits where the allowed singularities are fully fixed by conformal invariance.

In  AdS/CFT, sharp locality was analyzed in a seminal work by Heemskerk, Penedones, Polchinski, and Sully (HPPS) \cite{Heemskerk_2009}. {They demonstrated that the consistency of the CFT four point function, together with large-$N$ factorization and conformal symmetry, imposes stringent constraints on the allowed analytic structure of the correlators. In particular, 
the controlled behavior of logarithmic terms in specific limits imply that the dual bulk theory must be local up to a finite number of derivatives at each order in perturbation theory. In this sense, HPPS showed that locality in the bulk is not an independent assumption but an emergent consequence of CFT consistency conditions.}

Boundary conformal field theories (BCFTs) provide a clean and calculable setting to examine how introducing a boundary reshapes the conditions for bulk locality. The presence of a boundary reduces the symmetry group and modifies the kinematics of the correlation functions.  Bulk scalar primary operators acquire non-vanishing one point functions, and the two-point function of  scalar primaries depends nontrivially on a single conformal cross ratio \cite{McAvity:1995zd,Cardy:2004hm, Liendo:2012hy} (see e.g. \cite{Gliozzi:2015qsa, Billo:2016cpy, Herzog:2017xha, Andrei:2018die, Bissi:2018mcq, Lemos:2017vnx, Dey:2020lwp, Dey:2020jlc, Giombi:2020rmc,Lauria:2020emq, Bianchi:2022ppi, Chen:2023oax,  Giombi:2025pxx} for related works on CFTs with boundaries and defects). From the holographic point of view, BCFTs are described by AdS spacetimes terminated by an end-of-the-world (EOW) brane, as proposed in the AdS/BCFT construction \cite{Takayanagi:2011zk, Fujita:2011fp}. Boundary conditions imposed on the brane encode the BCFT data, while bulk fields propagate in a geometry that is globally distinct from pure AdS.

The BCFT two point function plays a role  analogous to that of the four point function in ordinary homogeneous CFT. {Its conformal block decomposition organizes contributions from bulk-channel and boundary-channel operators \cite{Liendo:2012hy}, and its logarithmic terms encode anomalous dimensions associated with bulk interactions. This makes the BCFT two point function a clean `observable' for probing sharp locality in the presence of boundaries.} From this perspective, one may regard BCFT two point functions as providing a lower-point, yet equally powerful, analogue of the HPPS four point function analysis.

Despite this close analogy, sharp locality in holographic BCFTs has received comparatively little systematic attention. The structure of BCFT correlators and their holographic representation have been explored from several complementary perspectives, including geodesic Witten diagram approximation \cite{Karch:2017wgy,Kastikainen:2021ybu} {and  identifying constraints on when a BCFT admits a semiclassical bulk brane description \cite{Reeves:2021sab}}. AdS correlators at finite temperature have been studied holographically in \cite{Alday:2020eua}. Since boundaries modify the global structure of spacetime and constrain bulk propagation, it is far from obvious that a local bulk EFT should remain consistent once quantum corrections are included. Whether the HPPS logic extends to BCFTs, or whether new phenomena appear, is therefore an important open question.

{In this work we address this question by performing an explicit  test of sharp locality in holographic BCFT in $d$ dimensions. We consider a scalar field propagating in AdS$_{d+1}$ with a planar EOW brane and study one-loop Witten diagrams generated by cubic and quartic bulk interactions. Using the method of images to implement boundary conditions, we compute the  contributions of these diagrams to the BCFT two-point function. This allows us to extract the analytic structure of the correlator and to compare it directly with the constraints imposed by the BCFT conformal block decomposition.}

Our analysis reveals a nontrivial and intriguing outcome. For generic values of the operator dimension $\Delta_{\phi}$, the loop-corrected two-point function derived from a local bulk action fails to match the analytic structure required by boundary conformal symmetry under standard Dirichlet or Neumann boundary conditions. Since these constraints follow purely from BCFT kinematics, independently of any bulk assumptions, this incompatibility demonstrates that a strictly local bulk EFT cannot consistently reproduce BCFT correlators beyond tree level. In this sense, our result provides a BCFT analogue of the HPPS locality test, but with a qualitatively different conclusion: in the presence of an EOW brane, sharp locality fails already at one-loop.

It is important to clarify the scope of this result. We do not interpret our findings as signaling a breakdown of the AdS/BCFT correspondence itself. Rather, our findings indicate that, under the assumptions of a purely local bulk effective field theory with standard boundary conditions, loop-level bulk amplitudes are not fully compatible with the analytic structure required by BCFT correlators. This suggests that additional ingredients -- such as boundary-localized interactions (see e.g. \cite{Kaviraj:2018tfd, Mazac:2018biw}), explicit defect degrees of freedom, or more general bulk descriptions beyond a local EFT -- may be necessary to reconcile loop-level bulk dynamics with BCFT analyticity. Understanding which of these possibilities is realized, and how consistency is restored, is an important step toward a more complete understanding of holography in the presence of boundaries.

More broadly, our work highlights BCFT correlators as a powerful and underutilized probe of microscopic bulk locality. Compared to four-point functions in ordinary CFTs, BCFT two-point functions are technically simpler while retaining sensitivity to loop effects and analytic constraints. This makes them an attractive arena for exploring how global geometric features, such as boundaries and defects, modify the emergence of spacetime and locality in holographic duals.

The rest of the paper is organized as follows. In Section \ref{bcftreview} we review some basics of the BCFT ideas. We discuss the computation of the two point scalar bulk correlators using Witten diagrams in Section \ref{bulkcomp}. Then we discuss the {incompatibility in analytic structure} of these Witten diagrams with the BCFT correlators. We conclude in Section \ref{concl} with discussions on our results and future directions. The appendices provide details on the Witten diagram computations.

\section{Review of boundary conformal field theory}\label{bcftreview}
In this section, we will review some basic concepts of a BCFT \cite{McAvity:1995zd, Cardy:2004hm, Liendo:2012hy}. We consider a conformal field theory in $d$-dimensions bounded by a flat surface at $\xp=0$. This is a BCFT defined in a semi-infinite space 
\begin{align}
x^\mu=(\Vec{x},\xp \geq 0)\ .
\end{align}
The boundary breaks translational invariance along $\xp$ direction. This BCFT can have two types of operators: bulk operators $\f(\xv,\xp)$, living on the entire space $x^\mu$ and boundary operators $\hat{\f}(\xv)$, localised on the boundary $\xp=0$. The observables we are interested in are the correlation functions of local bulk operators. The presence of the boundary results in a non-vanishing one-point function of bulk scalar operators. Hence the non-trivial dynamical information of a BCFT is encoded in the two point correlator of these operators. Let us consider the correlator of two identical scalar operators $\f$ with scaling dimension  $\D_\f$. The operators satisfy the following boundary conditions
\begin{align}\label{bcft_bdy}
\text{Dirichlet}: \quad \phi(\Vec{x},\xp=0)&=0\ \,  \text{or}\nn
\text{Neumann}: \quad \partial_{\xp}\phi(\Vec{x},\xp=0)&=0\,.
\end{align}
The correlator $\langle \f(\Vec{x}_1,\xp_1) \f(\Vec{x}_2,\xp_2) \rangle $ can be written in terms of the cross ratio $\xi$ as follows
\begin{align}
\langle \f(\Vec{x}_1,\xp_1) \f(\Vec{x}_2,\xp_2)\rangle&=\frac{F(\xi)}{(4 \xp_1 \xp_2)^{\D_\f}} \,,
\end{align}
where
\begin{align}
    \xi=\frac{(\Vec{x}_1-\Vec{x}_2)^2+(\xp_1- \xp_2)^2}{4\xp_1 \xp_2}\,.
\end{align} 
The function $F(\xi)$ admits an expansion in bulk channel as %or boundary channel conformal blocks  as
\begin{align}
 F(\xi) &= \xi^{-\D_\f}\sum_{\D \geq 0}a_\D \xi^{\D/2} \, _2F_1\bigg(\frac{\D}{2},\frac{\D}{2};\D+1-\frac{d}{2};-{\xi}\bigg) \label{bulkope}\,,\\
\end{align}
where $\D$ is the scaling dimension of the bulk 
exchanged operator. In this setup the spinning intermediate operators do not appear in the OPE as they are not consistent with the boundary conformal symmetry. The coefficients $a_\D$ are known as the OPE coefficients.

We are interested in a BCFT  with a small parameter $\epsilon$, such that in an interacting theory the BCFT data can be expanded around the generalised free theory. This data receives corrections proportional to $\epsilon$. Let us consider the bulk data that has the following expansion
\begin{align}\label{bcftdata}
\D_n&=2\D_\f+2n+\epsilon \,\gamma_n+O(\epsilon^2)\,,\nn
a_\D &= a^{(0)}_n+\epsilon \,a^{(1)}_n+O(\epsilon^2)\,,\quad
 n =0,1,2,\cdots\,.
\end{align}

In the bulk channel the correlator contains double trace operators $\f\Box^n \f$ with scaling dimension $\D_n$.
Here $\gamma_n$  
denotes the bulk anomalous dimension of these double trace operators. The expansion of the  function $F(\xi)$ then takes the following form
\begin{align}\label{pert_corr}
F(\xi) &=F^{(0)}(\xi)+\epsilon \, F^{(1)}(\xi)+O(\epsilon^2)\,.
\end{align}
We are interested in the anomalous dimensions  appearing in the BCFT spectrum. The bulk anomalous dimensions can be extracted from the $\log \xi$ term of $F^{(1)}(\xi)$ as
\begin{align}\label{blocklog}
F^{(1)}(\xi)&= \log \xi\sum_{n} \xi^{n}\frac{1}{2} \langle{a^{(0)}_n} \g_{n}\rangle\, _2F_1(n+\D_\f,n+\D_\f;2\D_\f+1+2n-\frac{d}{2};-{\xi})+\cdots \ ,
\end{align}
where the ellipsis denotes the non logarithmic terms.
The bracket $\langle \rangle $ denotes sum over possible degenerate operators in the spectrum. We can now focus on the discontinuities of $F^{(1)}(\xi)$ in $\xi$. The discontinuity of any function $f(\xi)$ is defined as
\begin{align}\label{eqn:disc}
 \underset{\xi}{\Disc}\, f(\xi) \equiv \lim_{\a \to 0^+} f(\xi + i \a) - f(\xi - i \a) \,.
\end{align}
Note that in \eqref{blocklog} there is a hypergeometric function with argument $-\xi$, which has a branch cut at $\xi \in (-\infty, -1)$ and is analytic elsewhere. On the other hand, the $\log \xi$ term has a branch cut at $\xi < 0$. As a consequence, the discontinuity in the range $\xi \in (-1,0)$ originates only from the logarithms in \eqref{blocklog}, which is $\underset{\xi<0}{\Disc}\, \log \xi=2\pi i$. Hence we have
\begin{align}\label{eqn:discblocklog}
\underset{-1<\xi<0}{\Disc}F^{(1)}(\xi)&= \pi i \sum_{n} \xi^{n} \langle{a^{(0)}_n} \g_{n}\rangle\, _2F_1(n+\D_\f,n+\D_\f;2\D_\f+1+2n-\frac{d}{2};-{\xi})\,.
\end{align}

We expect that such logarithmic singularities should also arise from the holographic dual correlator with bulk local interactions as we will discuss in the next section. Our goal is to match the discontinuities of the bulk channel expansion \eqref{eqn:discblocklog} with the ones of the Witten diagram computations.

\section{Bulk perspective of BCFT }\label{bulkcomp}
In this section we calculate the two point correlation function of a BCFT  using the bulk dual living in AdS space with an end-of-the-world brane. We consider a bulk  scalar field $\Phi$ with cubic and quartic interactions, propagating in AdS$_{d+1}$ in presence of a brane extended along the radial direction.  The bulk action in Poincar\'e patch coordinates $x^{\mu}=(y,\xp, \Vec{x})$ is given by  
\begin{align}\label{action}
S&=\int_{\xp \geq 0}\dd \xp \,\int_{y\geq 0} {\dd y}\, \int_{\mathbb{R}^{d-1}} \dd^{d-1}\xv\,\sqrt{g} \bigg(\frac{1}{2}\nabla_\mu\Phi\nabla^\mu \Phi+m^2\Phi^2+V(\Phi)\bigg)\,,
\end{align}
with the following interaction potential
\begin{align}\label{potential}
V(\Phi)=\d_{\nu, 1}{\left(\lambda_3 \F^3+\lambda_4 \Phi^4\right)}+ \d_{\nu, -1}\left(\lambda_5 (\nabla \Phi)^4+\cdots\,\right) .
\end{align}
The couplings are denoted by $\lambda_3, \lambda_4, \lambda_5$ and ellipsis denotes the higher derivative interactions. The $\d_{\nu,\pm1}$ serves a purpose that will become clear shortly. 
Here $g$ is the Euclidean AdS metric given by
\begin{align}
\dd s^2=\frac{1}{y^2}\left(\dd y^2+\dd{\xp}^2+\dd \Vec{x}^2\right)\,.
\end{align}

The bulk scalars can satisfy Dirichlet or Neumann boundary conditions\footnote{We impose Dirichlet or Neumann boundary conditions on the end-of-the-world brane, which ensure a well-defined variational principle.}
\begin{align}
\text{Dirichlet}: \quad \Phi(\Vec{x},\xp=0,y)&=0\ ,\nn
\text{Neumann}: \quad \partial_{\xp}\Phi(\Vec{x},\xp=0,y)&=0\,.
\end{align}
We use $\nu=+1$ for Neumann and $\nu=-1$ for Dirichlet boundary conditions respectively. Note that the Kronecker delta's in \eqref{potential} take care of the vertices relevant to a particular boundary condition. For example the first few simplest vertices on the brane are $\lambda_3 \F^3$ for Neumann and $\lambda_5 (\nabla \F)^4$ for Dirichlet.
The BCFT is located at the boundary $y=0$ and described by $(\xp\geq 0, \Vec{x})$.
The bulk field $\Phi$ can be thought of as dual to a scalar primary operator $\f$ in a BCFT, with scaling dimension $\D_\f$ satisfying 
\begin{align}
\D_\f (\D_\f-d)=m^2\,.
\end{align}

Using the holographic dictionary the boundary field $\f$ is defined by the limit 
\begin{align}
\Phi(\Vec{x},\xp,y) & \overset{y\rightarrow 0}\approx y^{\D_\f}\f(\Vec{x},\xp)\,,
\end{align}
and satisfies the same boundary conditions induced by the bulk AdS space \eqref{bcft_bdy}. 

Now we ask the question: Can we use the bulk action \eqref{potential} to compute the BCFT correlators with specific boundary conditions mentioned in Sec. \ref{bcftreview} and show their equivalence explicitly?

\subsection{Witten diagrams}
In this section we compute $\langle \f(\Vec{x}_1,\xp_1) \f(\Vec{x}_2,\xp_2)\rangle$ from the bulk action \eqref{action} using the Witten diagrams describing these correlators. In order to calculate the Witten diagrams, it is useful to use the embedding space formalism \cite{Weinberg:2012mz, Penedones:2007ns}. 
The embedding space coordinates are related to Poincar\'e patch coordinates as
\begin{align}\label{poincare}
P^A& =\{1, \Vec{x}^2+{\xp}^2, x^i,\xp\}\ ,\nn
X^A& =\frac{1}{y}\{1, \Vec{x}^2+{\xp}^2+y^2, x^i,\xp\}\ ,
\end{align}
where $i=1,2,\cdots d-1$ and $A=1,2,\cdots ,d+2$. The $d+2$-dimensional  vectors $P^A, X^A$ satisfy $X^2=-1, P^2=0$. 
In embedding space coordinates  the bulk-to-bulk and the bulk-to-boundary propagators in the absence of the brane read 
\begin{align}\label{propexp}
G_{B\partial}(P,X)&= \frac{1}{(-2P \cdot X)^{\D_\f} }\ ,\nn
G_{BB}(X_1,X_2) &= \frac{1}{\left(-{\zeta}\right)^{\D_\f}}\,  _2F_1\left(\D_\f,\D_\f-\frac{d-1}{2},2\D_\f-d+1,-\frac{4}{\zeta}\right)\ ,
\end{align}
where
\begin{align}\label{zetadef}
\zeta = \frac{(X_1-X_2)^2}{4} \ .
\end{align}
The propagators in \eqref{action} can be constructed from \eqref{propexp} via the method of images \cite{Kaviraj:2018tfd, Mazac:2018biw}
\begin{align}\label{propbcft}
G^{\nu}_{BB}(X_1,X_2)& =G_{BB}(X_1,X_2)+\nu \ G_{BB}(X_{1r},X_2)\,,\nn
G^{\nu}_{B\partial}(P,X) &=G_{B\partial}(P,X)+\nu \ G_{B\partial}(P,X_r)\,,
\end{align}
where $X_r= X \big{|}_{\xp\rightarrow -\xp}$\, . 

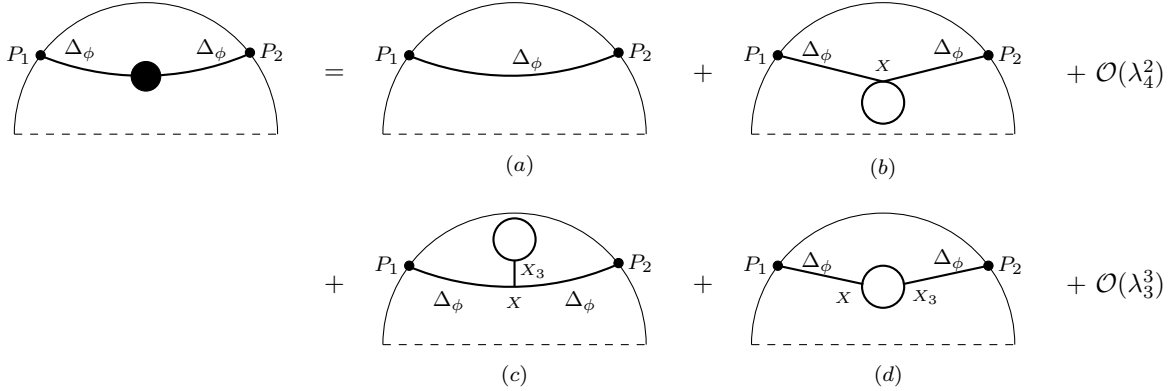
\begin{figure}[h]
 	{\small
 		\begin{center}
 			\begin{tikzpicture}[scale=0.35]
 				
 				\draw (0-14-14,0) arc (180:0:5) ;%-- (2,0);
 				\draw[dashed](0-14-14,0) -- (10-14-14,0);
 				%\draw[thick] (5-13,2) --(9-13,3);
 				%\draw[thick] (5-13,2) --(1-13,3);
 				\filldraw[black] (9.02-14.0-14.07,3.1) circle (5pt) node[anchor=west]{\scriptsize{$P_2$}};
 				\filldraw[black] (1-14-14,3) circle (5pt) node[anchor=east]{\scriptsize{$P_1$}};
 				\node at (2.5-14-14,3.2) {\scriptsize{$\Delta_\phi$}};
 				\node at (7.5-14-14,3.2) {\scriptsize{$\Delta_\phi$}};
 				\filldraw[black] (5-14-14,2.2) circle (16pt); %node[anchor=south]{$X$};
 				%\draw (5-1,1.2) circle (0.8);
 				
 				\draw[thick] (-4.2-14-1,3.) arc[start angle=-67.5, end angle=-112.8, radius=10.25];
 				
 				\node at (0-14.5-1, 2.3) {$=\ \ $};
 				
 				\draw (0,0) arc (180:0:5) ;%-- (2,0);
 				\draw[dashed](0,0) -- (10,0);
 				\draw[thick] (5,2) --(9,3);
 				\draw[thick] (5,2) --(1,3);
 				\filldraw[black] (9,3) circle (5pt) node[anchor=west]{\scriptsize{$P_2$}};
 				\filldraw[black] (1,3) circle (5pt) node[anchor=east]{\scriptsize{$P_1$}};
 				\node at (2.5,3.1) {\scriptsize{$\Delta_\phi$}};
 				\node at (7.5,3.2) {\scriptsize{$\Delta_\phi$}};
 				%\filldraw[black] (5,2) circle (3pt) node[anchor=south]{\scriptsize{$X$}};
 				\draw (5,1.2) circle (0.8);

 				\node at (0-1.5,2.3) {$+\ \ $};

 				\draw (0-14,0) arc (180:0:5) ;%-- (2,0);
 				\draw[dashed](0-14,0) -- (10-14,0);
 				%\draw[thick] (5-13,2) --(9-13,3);
 				%\draw[thick] (5-13,2) --(1-13,3);
               
 				\filldraw[black] (9.02-14.07,3.1) circle (5pt) node[anchor=west]{\scriptsize{$P_2$}};
 				\filldraw[black] (1-14,3) circle (5pt) node[anchor=east]{\scriptsize{$P_1$}};
 				\node at (2.5-11,3.2-0.4) {\scriptsize{$\Delta_\phi$}};
 				%\node at (7.5,3) {$\D_\f$};
 				% \filldraw[black] (5,2) circle (3pt) node[anchor=south]{$X$};
 				\draw[thick] (5,1.2) circle (0.8);
 				
 				\draw[thick] (-4.2-1,3.) arc[start angle=-67.5, end angle=-112.8, radius=10.25];

                \node at (2.5-11-.3,3.2-0.4-4) {\scriptsize{$(a)$}};

                \node at (14-8-1,2.6) {\tiny{$X$}};
                
 				\node at (14,2.3) {$ +\ \mathcal{O}(\lambda_4^2) \ $};

                \node at (14-8-1,- 1.3) {\scriptsize{$(b)$}};
 				
 				%5 The cubic contributions
 				\node at (0-15.5,2.3 - 8) {$+\ \ $};
 		    	\draw (0,-8) arc (180:0:5) ;%-- (2,0);
 				\draw[dashed](0,-8) -- (10,-8);
 				\draw[thick] (5+0.75,2-8+0.3) --(9,3-8);
 				\draw[thick] (5-0.75,2-8+0.3) --(1,3-8);
 				\filldraw[black] (9,3-8) circle (5pt) node[anchor=west]{\scriptsize{$P_2$}};
 				\filldraw[black] (1,3-8) circle (5pt) node[anchor=east]{\scriptsize{$P_1$}};
 				\node at (2.5,3.1-8) {\scriptsize{$\Delta_\phi$}};
 				\node at (7.5,3.2-8) {\scriptsize{$\Delta_\phi$}};
 				%\filldraw[black] (5,2) circle (3pt) node[anchor=south]{\scriptsize{$X$}};
 				\draw [thick] (5,1.2-8+1) circle (0.8);

                 \node at (14-10.5,2.3-8-.5) {\tiny $X$};
                
                \node at (14-8+0.6,2.3 -8-.5) {\tiny $X_3$};
 				
 				\node at (0-1.5,2.3 - 8) {$+\ \ $};

                \node at (14-8+0.6-1.4,2.3 -8-.5-3) {\scriptsize$(d)$};

 				\draw (0-14, - 8) arc (180:0:5) ;%-- (2,0);
 				\draw[dashed](0-14,-8) -- (10-14,-8);
 				%\draw[thick] (5-13,2) --(9-13,3);
 				%\draw[thick] (5-13,2) --(1-13,3);
                % \filldraw[black] (9.02-14.07,3.1) circle (5pt) node[anchor=west]{\scriptsize{$P_2$}};
 				\filldraw[black] (9.02-14.07,3.1-8) circle (5pt) node[anchor=west]{\scriptsize{$P_2$}};
 				\filldraw[black] (1-14,3-8) circle (5pt) node[anchor=east]{\scriptsize{$P_1$}};
 				\node at (2.5-14,3.2-0.4-7.5-1.7) {\scriptsize{$\Delta_\phi$}};
                \node at (2.5-9,3.2-0.4-7.5-1.7) {\scriptsize{$\Delta_\phi$}};
                
 				%\node at (7.5,3) {$\D_\f$};
 				% \filldraw[black] (5,2) circle (3pt) node[anchor=south]{$X$};

                 \node at (-8-0.3,-5.2) {\tiny $X_3$};
 				
 				\node at (-8-0.3-0.7,-5.2-4) {\scriptsize $(c)$};
                
 				\draw[thick] (-4.2-1,3.-8) arc[start angle=-67, end angle=-113, radius=10.1];
 				
 				\draw[thick] (5-14,1.2-5.2) circle (0.8);
 				\draw[thick](0-14+5,-5.8) -- (0-14+5,-4.8);

               \node at (2.5-11-0.5,3.2-0.4-9.2) {\tiny{$X$}};

 				\node at (14,2.3 -8) {$ +\ \mathcal{O}(\lambda_3^3) \ $};
 				
 			\end{tikzpicture}
 			
 			\caption{$\langle \f(P_1)\f(P_2)\rangle $ in the background \eqref{potential} due to $\lambda_3 \Phi^3  +\lambda_4 \Phi^4$ interactions}
 			\label{fig:bulk-propagator-corrections}
 	\end{center}}
 \end{figure}

With these propagators, we will now compute the leading corrections to the two point correlator $\langle \phi (P_1)\phi(P_2)\rangle $ for different boundary conditions, perturbatively in the couplings  in \eqref{potential}. For simplicity we use Neumann boundary conditions for $\F$. This can be generalised to Dirichlet boundary conditions. The leading  Witten diagrams corresponding to the Neumann boundary conditions are shown in Figure \ref{fig:bulk-propagator-corrections}. 

To compute the Witten  diagrams we use the Schwinger parameter representation of the bulk-to-boundary propagators 
\begin{align}\label{bulk_bdy1}
    G_{B\partial}(P_i,X)
    &=\frac{y^{\D_\f}}{\Gamma(\D_\f)}\int_0^{\infty} \dd s_i \, {s_i}^{\D_\f-1} \, e^{-s_i\left(y^2+(x_i^{\perp}-\xp)^2+(\Vec{x}_i-\xv)^2\right)}\,,
\end{align}
where 
\begin{align}
-2P_i\cdot X & =\frac{1}{y}\left(y^2+(x_i^{\perp}-\xp)^2+(\Vec{x}_i-\xv)^2\right)\ .
\end{align}
We also use the Mellin-Barnes representation of the Gaussian hypergeometric function for the bulk-to-bulk propagator 
\begin{align}\label{bulk_bulk_1}
G_{BB}(X_1,X_2)
&=\frac{(-1)^{\D_\f}\Gamma(2\D_\f-d+1)}{\Gamma(\D_\f)\Gamma(\D_\f-\frac{d-1}{2})}\int_{s_0-i \infty}^{s_0+i \infty}  \frac{\dd s}{2\pi i}\frac{\Gamma(s)\Gamma(\D_\f-s)\Gamma(\D_\f-\frac{d-1}{2}-s)}{\Gamma(2\D_\f-d+1-s)} {4^{-s}}{\zeta^{s-\D_\f}}\, ,
\end{align}
where $\zeta$ is defined in \eqref{zetadef}. 

The integration contour is located at $0< s_0< \text{min}(\text{Re}\, a,\text{Re}\, b)$ i.e. $0< s_0< \D_\f-\frac{d-1}{2}$. Note that in Figure \ref{fig:bulk-propagator-corrections} there are bulk-to-bulk propagators between coincident points. This implies a divergence coming from  $G_{BB}(X_1,X)$ as $X_1 \rightarrow X$. We assume such terms can be renormalised and we set these terms to zero. Hence we have

\begin{align}\label{renorm}
\lim_{X_1 \rightarrow X } G^{\nu}_{BB}(X_1,X)& =\nu\,  G_{BB}(X_{r},X)\ .
\end{align}

We now have all the necessary ingredients to compute the Witten diagrams.
In the following subsections we evaluate 
the Witten diagrams in Figure \ref{fig:bulk-propagator-corrections} due to $\lambda_3 \F^3$ and $\lambda_4\F^4$ and compare it with the BCFT correlator \eqref{pert_corr}. Specifically we want to focus on the corrections of the correlator related to the bulk anomalous dimensions of the BCFT data \eqref{blocklog} in what follows. {{We assume that any local theory in the bulk can be derived from a local Lagrangian in the weak coupling expansion in $\lambda_3$ and $\lambda_4$. Our goal is to see if the structure of the BCFT correlator is compatible with the correlator derived from the bulk Lagrangian \eqref{action}.}

\subsection{Quartic vertex}

Let us now compute the leading correction to  $\langle \f(P_1)\phi(P_2)\rangle$ due to $\lambda_4 \Phi^4$ interaction in \eqref{action}. The one-loop Witten diagram is given by the diagram depicted in Figure \ref{fig:bulk-propagator-corrections}(b) 
\begin{align}\label{2ptfn}
\langle \f(P_1) \f(P_2)\rangle \bigg|_{\lambda_4}%&=\int dX \sqrt{g}\,\langle \F(P_1) \F(X)\rangle \langle \F(P_2) \F(X)\rangle \langle \F(X) \F(X)\rangle \nn
&= \int \dd X \sqrt{g}\, G^{\nu}_{B\partial}(P_1,X) G^{\nu}_{B\partial}(P_2,X) G^{\nu}_{BB}(X,X)\,,\nn
%&=\nu \int \dd X \sqrt{g}\, G^{\nu}_{B\partial}(P_1,X) G^{\nu}_{B\partial}(P_2,X) G_{BB}(X_r,X)\,. 
&=  \int \dd X \sqrt{g}\,\bigg( G_{B\partial}(P_1,X) G_{B\partial}(P_2,X)+ G_{B\partial}(P_1,X_r) G_{B\partial}(P_2,X)\nn &+G_{B\partial}(P_1,X) G_{B\partial}(P_2,X_r)+ G_{B\partial}(P_1,X_r) G_{B\partial}(P_2,X_r)\bigg) G_{BB}(X_r,X)\,.
\end{align}
Here we have used \eqref{propbcft} to get the second line of \eqref{2ptfn}. The volume form is given by $\sqrt{g}=y^{-d-1}$. The integrals can be computed using the tools discussed in \cite{Rastelli:2017ecj}. Note that \eqref{2ptfn} contains four terms. We sketch below the details to evaluate the first term. The other terms can be evaluated using similar methods. Let us consider the term
\begin{align}\label{2pnt1}
\mathcal{C}&= \int \dd X \sqrt{g}\, G_{B\partial}(P_1,X) G_{B\partial}(P_2,X) G_{BB}(X_r,X)\,.
\end{align}
 We use  \eqref{bulk_bdy1} and \eqref{bulk_bulk_1} to rewrite the integral as
 \begin{align}
\mathcal{C}&=\mathcal{C}_0 \int_0^{\infty} \dd\xp\int_{\mathbb{R}^{d-1}} \dd^{d-1}x\int_0^{\infty}\dd y\ {y^{4\D_\f-2s-d-1}} \prod_{i=1}^2\bigg(\int_0^{\infty} \dd s_i \, {s_i}^{\D_\f-1}  e^{-s_i\left(y^2+(x_i^{\perp}-\xp)^2+(\Vec{x}_i-\xv)^2\right)}\bigg)\nn & \times \int_{s_0-i \infty}^{s_0+i \infty}  \frac{\dd s}{2\pi i}\frac{\Gamma(s)\Gamma(\D_\f-s)\Gamma(\D_\f-\frac{d-1}{2}-s)}{\Gamma(2\D_\f-d+1-s)} {4^{-s}}{\xp}^{2(s-\D_\f)}\,,
 \end{align}
 where
 \begin{align}
\mathcal{C}_0 =\frac{(-1)^{\D_\f}\Gamma(2\D_\f-d+1)}{\Gamma^3(\D_\f)\Gamma(\D_\f-\frac{d-1}{2})}\,.
 \end{align}
 We first swap the contour integrals with the bulk coordinate integrals. Then we do the integrals over the bulk coordinates $y, \xv $ and $\xp$ one by one. The $y$ and $\xv$ integrals are straightforward to evaluate.
We quote below the results of these integrals
\begin{align}
   \int_0^{\infty} \dd y \,y^{4\D_\f-2s-d-1} \, e^{-(s_1+s_2)y^2}&= \frac{1}{2}(s_1+s_2)^{s+\frac{d}{2}-2\D_\f}\Gamma \left(-\frac{d}{2}+2 \D_\f-s\right)\,,
\end{align}
\begin{align}
    \int_{\mathbb{R}^{d-1}} \dd^{d-1}\xv\ e^{-s_1(\Vec{x}_1-\xv)^2-s_2(\Vec{x}_2-\xv)^2}=\left(\frac{\pi}{s_1+s_2}\right)^{\frac{d-1}{2}}e^{-\frac{s_1 s_2}{s_1+s_2}(\Vec{x}_1-\Vec{x}_2)^2}\,.
\end{align}
The $\xp$ integral can be evaluated as follows. We first use the Mellin representation of $e^{-z}$ 
\begin{align}\label{expint}
e^{-z}=\int_{- i \infty}^{i \infty} \frac{\dd \tau }{2\pi i}\,\Gamma(\tau) \,  z^{-\tau}\,,
\end{align}
and rearrange the exponential as
\begin{align}
& e^{-\frac{s_1 s_2}{s_1+s_2}(\Vec{x}_1-\Vec{x}_2)^2}e^{-s_1(x_1^{\perp}-\xp)^2}e^{-s_2(x_2^{\perp}-\xp)^2}\nn %&=e^{-\frac{s_1 s_2}{s_1+s_2}(\Vec{x}_1-\Vec{x}_2)^2}e^{-\frac{s_1 s_2}{s_1+s_2}(x_1^{\perp}-x_2^{\perp})^2}e^{-(s_1+s_2)\left(\xp-\frac{s_1x_1^{\perp}+s_2 x_2^{\perp}}{s_1+s_2}\right)^2}\nn
&=e^{-\frac{s_1 s_2}{s_1+s_2}(4 \xp_1\xp_2\xi)}\int_{- i \infty}^{i \infty} \frac{\dd \tau}{2\pi i} \, \Gamma(\tau) (s_1+s_2)^{-\tau}\left(\xp-\frac{s_1x_1^{\perp}+s_2 x_2^{\perp}}{s_1+s_2}\right)^{-2\tau}\,.
\end{align}
This allows us to do the $\xp$ integral resulting in the following expression
\begin{align}
&\int_0^{\infty} \dd \xp \,  {\xp}^{2(s-\Delta_\phi)}\left(\xp-\frac{s_1x_1^{\perp}+s_2 x_2^{\perp}}{s_1+s_2}\right)^{-2\tau}\nn &=(-1)^{1-2s+2\D_\f-2\tau}\frac{\Gamma(1+2s-2\D_\f)\G(2\D_\f+2\tau-2s-1)}{\Gamma(2\tau)}\left(\frac{s_1+s_2}{s_1 x_1^{\perp}+s_2 x_2^{\perp}}\right)^{2\D_\f+2\tau-2s-1} \,.
\end{align}
Now we are left with the $s_1,s_2, s, \tau$ integrals. The $s_1,s_2$ integrals can be done by introducing a new variable $\rho$ such that
\begin{align}
1=\int_0^{\infty} \dd \rho\, \d(\rho-s_1-s_2)\,.
\end{align}
At this level we rescale $s_i\rightarrow \rho \ {s}_i\,, i=1,2$ and use the following identity
\begin{align}
\d(\rho-\rho {s}_1-\rho{s}_2)=\frac{1}{\rho}\, \d(1-{s}_1-{s}_2)\,.
\end{align}
Note that the delta function $\d(1-s_1-s_2)$ restricts the integration region of $s_1, s_2$ to $[0,1]$. The $\rho$ integral can be done first and then the $s_1, s_2$-integrals using the Feynman parameterization formula
\begin{align}
\frac{1}{A_1^{\alpha_1}A_2^{\alpha_2}}=\frac{\G(\a_1+\a_2)}{\G(\a_1)\G(\a_2)}\int_0^1\int_0^1 \dd s_1 \dd s_2 \,s^{\a_1-1}_1s^{\a_2-1}_2\frac{\d(1-s_1-s_2)}{(s_1 A_1+s_2 A_2)^{\a_1+\a_2}}\, .
\end{align}
We are left with the following 
\begin{align}
\mathcal{C} &=
\mathcal{C}_0\frac{(-1)^{1+2\D_\f}{\pi^{\frac{d}{2}}}}{2(x_1^{\perp}x_2^{\perp})^{\D_\f}}\int_{s_0-i \infty}^{s_0+i \infty} \frac{\dd s}{2\pi i} \frac{(-1)^{-2s}\G(s)\G(\D_\f-s)\G(\D_\f-\frac{d-1}{2}-s) }{2^{4s}\ \G(2\D_\f-d+1-s)}\G(1+2s-2\D_\f) \nn & \times\G\left(2\D_\f-s-\frac{d}{2}\right) \int_{- i \infty}^{i \infty} \frac{\dd\tau}{2\pi i}\ (-1)^{-2\tau} \xi^{\tau-s-\frac{1}{2}}\frac{\G(s-\tau+\frac{1}{2})\G^2(\D_\f+\tau-s-\frac{1}{2})}{\G(\tau+\frac{1}{2})} \,.
\end{align}
The $\tau$ integral results in
\begin{align}
 \int_{- i \infty}^{i \infty} \frac{\dd\tau}{2\pi i}\ (-1)^{-2\tau} \xi^{\tau-s-\frac{1}{2}}\frac{\G(s-\tau+\frac{1}{2})\G^2(\D_\f+\tau-s-\frac{1}{2})}{\G(\tau+\frac{1}{2})}=\frac{\G^2(\D_\f)}{\G(s+1)} \,  _2F_1(\Delta_\phi ,\Delta_\phi ;s+1;-\xi )\,.
\end{align}
Hence we have
\begin{align}
\mathcal{C} &=
\mathcal{C}_0\frac{(-1)^{1+2\D_\f}{\pi^{\frac{d}{2}}}\G^2(\D_\f)}{2(x_1^{\perp}x_2^{\perp})^{\D_\f}}\int_{s_0-i \infty}^{s_0+i \infty} \frac{\dd s}{2\pi i}\frac{(-1)^{-2s}}{2^{4s}} \frac{\G(1+2s-2\D_\f)\G(s)\G(\D_\f-s)\G(\D_\f-\frac{d-1}{2}-s)}{\G(s+1)\G(2\D_\f-d+1-s)}\nn & \times {\G\left(2\D_\f-s-\frac{d}{2}\right)}\,  _2F_1(\Delta_\phi ,\Delta_\phi ;s+1;-\xi )  \,.
\end{align}

We evaluate the remaining terms of \eqref{2ptfn} using similar techniques. Finally, collecting all the terms we have
\begin{align}\label{eqn:res:2}
\langle \f(P_1) \f(P_2)\rangle \bigg|_{\lambda_4} &=\mathcal{D}_0\int_{s_0-i \infty}^{s_0+i \infty} \frac{\dd s }{2\pi i}\,\mathcal{A}(s) \bigg( [1-(-1)^{2s-2\D_\f}]\mathcal{F}_s(\xi)+2(-1)^{4s-3\D_\f}\mathcal{F}_s({-1-\xi})\bigg)\,,
\end{align}
where
\begin{align}
\mathcal{D}_0&= \frac{ (-1)^{3\D_\f-1}}{(\xp_1 \xp_2)^{\D_\f}}\frac{\G(\D_\f+1-\frac{d}{2})}{2^{d-2\D_\f+1}\pi^{\frac{1-d}{2}}\G(\D_\f)}\,,\nn
\mathcal{A}(s)&= (-1)^{-2s}\frac{\Gamma(\D_\f-s)\Gamma(\D_\f-s-\frac{d-1}{2})}{2^{4s}s\Gamma(1-d-s+2\D_\f)}\Gamma(1+2s-2\D_\f)\Gamma(2\D_\f-s-\frac{d}{2})\,,\label{poles}\\
\mathcal{F}_s(\xi)&=\, _2F_1(\Delta_\phi ,\Delta_\phi ;s+1;-\xi )\,.\label{eqn:fxi}
\end{align}
Now we evaluate the $s$-integral in \eqref{eqn:res:2} using residue theorem 
by closing the $s$-contour to the left $s$-plane. This will have the following contributions from the poles  in \eqref{poles} at 
\begin{enumerate}[(i)]
    \item $s=\frac{1}{2}(2\D_\f-r-1)\,,\quad r \in \mathbb{Z}_{\geq 0}$ from  $\Gamma(1+2s-2\D_\f)$ \,. 
    \item simple pole at $s=0$\, .
\end{enumerate}
This results in the following 
\begin{align}\label{eqn:resfinal}
&\langle \f(P_1) \f(P_2)\rangle \bigg|_{\lambda_4}\nn &=\mathcal{D}_0 \frac{\sqrt{\pi}e^{-3i \pi \D_\f}\Gamma (1-2 \Delta_\phi ) \Gamma (\Delta_\phi ) \Gamma \left(2 \Delta_\phi -\frac{d}{2}\right)}{2^{2\D_\f-d}\G(1+\D_\f-\frac{d}{2})}  \bigg([(-1)^{3\D_\f}-e^{i \pi \D_\f}]\mathcal{F}_0(\xi)+2\mathcal{F}_0({-1-\xi})\bigg)\nn
&+\mathcal{D}_0\sum_{r=0}^{\infty} \frac{\sqrt{\pi } (-1)^{r-2 \Delta_\phi }  \Gamma \left(\frac{1}{2} (-d+r+2)\right) \Gamma \left(\frac{1}{2} (-d+r+1)+\Delta_\phi \right)}{ 2^{4 \Delta_\phi -r-2} (-2 \Delta_\phi +r+1) \Gamma \left(\frac{r}{2}+1\right) \Gamma \left(-d+\frac{r+3}{2}+\Delta_\phi \right)}\nn &\times\bigg([1 + (-1)^r]\mathcal{F}_{\D_\f-\frac{r+1}{2}}(\xi)+2e^{i\pi (\D_\f+r)}\mathcal{F}_{\D_\f-\frac{r+1}{2}}({-1-\xi})\bigg)\,.
\end{align}
This completes the evaluation of the diagram Figure \ref{fig:bulk-propagator-corrections}(b). One can read-off the coefficients of $\log \xi$ from \eqref{eqn:resfinal} which, in turn, provides us with the information of the anomalous dimensions of the bulk operators in the BCFT as mentioned in \eqref{blocklog}. However, expanding \eqref{eqn:resfinal} around $\xi\sim 0$ does not give any $\log \xi$ term, but only power law terms in $\xi$.

At this level we can further use \eqref{eqn:disc}  to calculate the discontinuity of \eqref{eqn:resfinal}. Note that the Witten diagram contains two hypergeometric functions with arguments $-\xi$ and $1+\xi$, which have branch cuts at $\xi \in (-\infty, -1)$ and $\xi \in (0,\infty)$ respectively . As a result, the discontinuity of \eqref{eqn:resfinal} in the range $\xi \in (-1,0)$ vanishes. Matching the discontinuity of \eqref{eqn:resfinal} with \eqref{eqn:discblocklog} we get
\begin{align}\label{eqn:anmdimvanish}
\langle a^{(0)}_n\g_n\rangle =0\,, \forall n\,.
\end{align}
The statement \eqref{eqn:anmdimvanish} suggests a mismatch in the analytic structure of the correlator obtained via Witten diagrams and BCFT. This incompatibility indicates that a purely local quartic bulk interaction is insufficient to reproduce BCFT correlator at loop level under standard boundary conditions.

{It is therefore natural to investigate whether one-loop diagrams generated by cubic bulk interactions can alter this conclusion, and in particular whether their contributions can restore the analytic structure required by boundary conformal symmetry. This possibility is analyzed in the subsequent subsection.
}

\subsection{Cubic vertex}
In this section we calculate the leading correction of $\langle \f(P_1)\f(P_2)\rangle$ due to cubic interaction $\lambda_3 \Phi^3$  in \eqref{action}. The leading correction comes from  the Witten diagrams at $O(\lambda^2_3)$ as depicted in  (c) and (d) of Figure \ref{fig:bulk-propagator-corrections}. These are the tadpole and bubble diagrams which can be evaluated by the following
\begin{align}
\langle \f(P_1) \f(P_2)\rangle \bigg|_{\lambda_3^2}&=\mathcal{I}_{tad}+\mathcal{I}_{bub}\,, \nn 
\text{where}\,\quad \mathcal{I}_{tad} &= \int \dd X_3 \sqrt{g} \int \dd X \sqrt{g}\ G^{\nu}_{B\partial}(P_1,X) G^{\nu}_{B\partial}(P_2,X) G^{\nu}_{BB}(X_,X_3) G^{\nu}_{BB}(X_{3r},X_3)\,, \label{tad1}\\
\mathcal{I}_{bub} &=  \int \dd X_3 \sqrt{g} \int \dd X \sqrt{g}\ G^{\nu}_{B\partial}(P_1,X) G^{\nu}_{B\partial}(P_2,X_3) G^{\nu}_{BB}(X_,X_3)^2\, \label{bub1}.
\end{align}
These integrals can be evaluated using the methods described in the previous subsection. The details can be found in the appendices \ref{tadpole} and \ref{bubble}. It turns out that there are no contribution coming from the bubble diagram. 
Taking into account the contributions from the tadpole diagram we get the following
\begin{align}\label{eqn:resfinaltad}
&\langle \f(P_1) \f(P_2)\rangle \bigg|_{\lambda^2_3}\nn &=\mathcal{P}_0\int \frac{\dd s}{2 \pi i}\int\frac{\dd s_4}{2\pi i}\int\frac{\dd \alpha}{2 \pi i}\,\int\frac{\dd \tau_1}{2 \pi i}\,\mathcal{B}(s,\tau_1,\alpha,s_4)\,\nn & \times \bigg( \big[(-1)^{-2s+2\D_\phi+2\tau_1}-(-1)^{2s-2\D_\f-2\tau_1}\big]\mathcal{F}_{s-\tau_1+\frac{1}{2}}(\xi)+2(-1)^{1+6s-5\D_\f-6\tau_1}\mathcal{F}_{s-\tau_1+\frac{1}{2}}({-1-\xi})\bigg)\,,
\end{align}
where 
\begin{align}
\mathcal{P}_0&= \frac{ (-1)^{2\D_\f}}{(\xp_1 \xp_2)^{\D_\f}}\frac{\G^2(\D_\f+1-\frac{d}{2})}{2^{2d-6\D_\f}\pi^{\frac{1}{2}-d}\G^2(\D_\f)}\,,\label{eqn:p0}\\
\mathcal{B}(s,\tau_1,\alpha,s_4)&= \frac{(-1)^{d+s_4-2\alpha-3\D_\phi}\,\G(s_4)\,\G(\D_\phi-s)\,\G(\D_\phi-\frac{d-1}{2}-s)\G(\D_\phi-\frac{d-1}{2}-s_4)}{2^{4s+2\alpha+4s_4}\,\G(2\D_\phi-d+1-s)\G(2\D_\phi-d+1-s_4)\,\G\big(\a+\frac{1}{2}\big)\,\G\big(\tau_1+\frac{1}{2}\big)\,\G\big(\frac{3}{2}+s-\tau_1\big)}\,\nn & \times \G(d+s_4+2\a-3\D_\f)\,\G(-d-s_4+3\D_\f)\,\G(-d-s-s_4-\a+3\D_\f)\,\nn & \times \G(1+2s-2\D_\f)\,\G(-1-2s+2\D_\f+2\tau_1)\,\G(2+2s-2\D_\f-2\tau_1)\nn & \times \G\bigg(\frac{1-d}{2}+s-s_4-\a+\D_\f-\tau_1\bigg)\,\G\bigg(\frac{d-1}{2}+s_4+\a-\D_\f+\tau_1\bigg)\,.\label{eqn:b} %,\nn
%\mathcal{F}_s(\xi)&=\, _2F_1(\Delta_\phi ,\Delta_\phi ;s+1;-\xi )\,.
\end{align}
Here also we find that the Witten diagram contains contributions from two hypergeometric functions with arguments $-\xi$ and $1+\xi$, which have branch cuts at $\xi \in (-\infty, -1)$ and $\xi \in (0,\infty)$ respectively. Hence the cubic diagrams can also not reproduce the analytic structure of the BCFT correlator \eqref{eqn:discblocklog}.

To summarize our findings, for generic values of $\D_\f$, the bulk interactions in \eqref{potential} could not generate the  terms at one-loop in the two point correlator of the BCFT which we expect from \eqref{blocklog}.  This incompatibility cannot be removed by boundary-localized counter terms or field redefinitions. Note that purely local brane terms, being analytic functions of the cross-ratio cannot generate the desired non-analytic branch cut in \eqref{blocklog}. We have demonstrated this for Neumann boundary conditions. This also holds for Dirichlet boundary condition with quartic interaction vertex.

\section{Discussions and conclusion}\label{concl}

We have shown that the one-loop correction to the BCFT two-point correlator obtained from a local bulk effective action in AdS with an end-of-the-world brane is not compatible with the analytic structure required by boundary conformal symmetry under standard boundary conditions. This places a sharp constraint on the validity of a strictly local bulk description.

Our results show that maintaining consistency at loop level requires additional ingredients beyond a local bulk effective field theory with standard boundary conditions. Possible resolutions include brane-localized interactions, explicit defect degrees of freedom, modified boundary conditions, or intrinsically nonlocal bulk dynamics. More broadly, this work establishes BCFT two-point functions as precise probes of microscopic bulk locality and provides a concrete framework for exploring how boundaries and defects modify the emergence of spacetime in holographic duals.

Our analysis has been carried out in a minimal scalar field setting in order to isolate the analytic structure of loop-level bulk correlators as cleanly as possible. We expect that this mechanism is not specific to scalars and should persist more generally, although its detailed manifestation may depend on the specific model. In more realistic constructions -- such as Wilson-line defects or particular D-brane arrangements -- additional degrees of freedom localized on the defect or brane are present and  play an essential role in the dynamics. It is not inconceivable that such additional degrees of freedom modify the correlators in a way that alters our conclusion. Understanding bulk locality in such  explicit top-down constructions is an important direction for future work.

There are several natural extensions of the present analysis. As indicated above, it would be interesting to understand whether the observed breakdown of sharp bulk locality persists for fields with spin, in particular for gauge fields and gravitons, where boundary conditions are more constrained by symmetry. Another important direction is to investigate whether the incompatibility can be resolved by introducing explicit boundary degrees of freedom or brane-localized interactions (see e.g. \cite{Kaviraj:2018tfd,Mazac:2018biw}), and how such modifications are encoded in BCFT data. Extending the analysis to higher-loop corrections, other geometries with defects, or to time-dependent and Lorentzian setups may further clarify the interplay between boundaries, quantum corrections and bulk locality. {It would also be interesting to include dynamical gravity and examine whether gravitational backreaction or induced gravitational dynamics on the brane modify the loop-level analytic structure of BCFT correlators. Moreover, semiclassical and geodesic approximations to the relevant Witten diagrams may provide useful geometric insight into the origin of the non-analytic structures and help connect exact loop results with an effective spacetime picture of bulk propagation.} Finally, it would be valuable to develop a systematic criterion -- analogous to the HPPS analysis in standard AdS/CFT -- for characterizing when holographic theories with boundaries admit a sharply local bulk description. {We hope to investigate some of these intriguing possibilities in future work.}

\acknowledgments{We thank Apratim Kaviraj and Alok Laddha for useful discussions. {We also thank the referee for constructive and helpful suggestions.} PB would like to acknowledge the support provided by 
Anusandhan National Research Foundation (ANRF), India through the Ramanujan fellowship grant RJF/2023/000007. PD is supported by ANRF Early Career Research Grant ANRF/ECRG/2024/000247/PMS. PD thanks the Yukawa Institute for Theoretical Physics at Kyoto University, where this work was completed during “Progress of Theoretical Bootstrap”.
}

\appendix
\section{Details of tadpole diagram}\label{tadpole}

In this section we discuss the details of evaluating the integral in \eqref{tad1}. There are total eight terms in the integrand. We discuss below the procedure to evaluate one of these terms. The other terms can be evaluated using similar techniques. Let us evaluate the following term for $\nu=1$
\begin{align}\label{tad11}
\mathcal{I}^{(1)}_{tad} &= \int \dd X_3 \sqrt{g} \int \dd X \sqrt{g}\ G_{B\partial}(P_1,X) G_{B\partial}(P_2,X) G_{BB}(X_,X_3) G_{BB}(X_{3r},X_3) \,.
\end{align}
Using \ref{bulk_bulk_1} we write
\begin{align}
 G_{BB}(X,X_3)
&=  (-1)^{\D_\f}\frac{\Gamma(c)}{\Gamma(a)\Gamma(b)}\int_{s_0-i \infty}^{s_0+i \infty}  \frac{\dd s_4}{2\pi i}\frac{\Gamma(s_4)\Gamma(a-s_4)\Gamma(b-s_4)}{\Gamma(c-s_4)} {4^{-s_4}}{\zeta^{s_4-\D_\f}_1}\,,
\end{align}
{with} 
\begin{align}
\zeta_1  &=\frac{-2-2X_3.X}{4} 
=\frac{\bar{\zeta}}{4 y \ y_3}\,,
\end{align}
{and}
\begin{align}
\bar{\zeta} &=(y-y_3)^2+(\xp-\xp_3)^2+(\xv-{\xv}_3)^2\,.
\end{align}
The $a,b,c$'s are defined as
\begin{align}
 a&=\D_\f,\quad b= \D_\f-\frac{d-1}{2},\quad  c=2\D_\f-d+1\,.
 \end{align}

Next we write $\bar{\zeta}^{s_4-\D_\f}$ in terms of Schwinger parameter
\begin{align}
{\bar{\zeta}^{s_4-\D_\f}}=\frac{1}{\Gamma(\D_\f-s_4)}\int_0^{\infty} \dd s_5 s^{\D_\f-s_4-1}_5\ e^{-s_5 \big((y-y_3)^2+(\xp-\xp_3)^2+(\xv-{\xv}_3)^2\big)}\,.
\end{align}
Putting these  together we write \eqref{tad11} as 
\begin{align}\label{tadrep}
\mathcal{I}^{(1)}_{tad}&=\int_0^{\infty} \dd s_5 s^{\D_\f-s_4-1}_5 \int  \frac{\dd s}{2\pi i} \frac{\dd s_4}{2\pi i}\frac{\Gamma(s)\Gamma(a-s)\Gamma(b-s)\Gamma(s_4)\Gamma(b-s_4)}{4^{s+2s_4-\D_\f}\Gamma(c-s)\Gamma(c-s_4)}  \bigg(\prod_{i=1}^2\int_0^{\infty} \dd s_i s_i^{\D_\f-1} \bigg)  \mathcal{I}_{tot}\,,
\end{align}
where
\begin{align}\label{itot}
\mathcal{I}_{tot}=&\int\dd y \,  \dd \xp \, \dd^{d-1}\xv \ \dd y_3\, \dd \xp_3\,  \dd^{d-1}\xv_3 \ (\xp_3)^{2s-2\D_\f}  y^{3\D_\f-d-1-s_4} y^{3\D_\f-d-1-s_4-2s}_3\nn
& \times e^{-s_5 \big((y-y_3)^2+(\xp-x_3^{\perp})^2+(\xv-{\xv}_3)^2\big)} e^{-s_1\big(y^2+(x_1^{\perp}-\xp)^2+(\Vec{x}_1-\xv)^2\big)}
 e^{-s_2\big(y^2+(x_2^{\perp}-\xp)^2+(\Vec{x}_2-\xv)^2\big)}\,.
\end{align}
We  perform the integrals over the bulk coordinates $y, \xp, \xv,y_3, \xp_3, \xv_3$ in \eqref{itot}. First we do the $y,y_3$ integrals by rewriting the exponentials as
\begin{align}
e^{-s_5 (y-y_3)^2} e^{-s_1 y^2} e^{-s_2 y^2 }=e^{-(s_1+s_2+s_5)\left(y-\frac{s_5 y_3}{s_1+s_2+s_5}\right)^2} e^{-y^2_3 \frac{s_5 (s_1+s_2)}{s_1+s_2+s_5}}\,.
\end{align}
Then using the integral representation  \eqref{expint} we write
\begin{align}
e^{-(s_1+s_2+s_5)\left(y-\frac{s_5 y_3}{s_1+s_2+s_5}\right)^2} =\int_{-i \infty}^{i \infty} \frac{\dd \alpha}{2\pi i}  \ \Gamma(\alpha)\ (s_1+s_2+s_5)^{-\alpha} \left(y-\frac{s_5\  y_3}{s_1+s_2+s_5}\right)^{-2\alpha} \,.
\end{align}
Now we do the $y$ and $y_3$ integrals which results in
{
\begin{align}
&\int_0^{\infty } \dd y \int_0^{\infty }  \dd y_3\, y^{-d-1+3\D_\f-s_4}{y_3^{3\D_\f-2s-d-1-s_4}}e^{-s_5 (y-y_3)^2-s_1 y^2-s_2 y^2 }\nn &= \int_{-i \infty}^{i \infty} \frac{\dd \alpha}{2\pi i} { (-1)^{-2 \alpha +d-3\Delta _{\phi }+s_4} \Gamma \left(-d-s_4+3\Delta _{\phi }\right) \Gamma \left(d+2 \alpha +s_4-3\Delta _{\phi }\right) \Gamma \left(-d-s-\alpha -s_4+3 \Delta _{\phi }\right)}  \nn
& \times \frac{\Gamma (\alpha )}{2 \Gamma (2 \alpha )} \left(s_1+s_2+s_5\right){}^{-s} s_5^{-\alpha +s} \left(s_1+s_2\right){}^{\alpha +d-3 \Delta _{\phi }+s+s_4}\,.
\end{align}}
Next we do the $\xv, {\xv}_3$ integrals
{
\begin{align}
\int_{\mathbb{R}^{d-1}} \dd^{d-1}\xv \ \dd^{d-1}\xv_3 \ e^{-s_5 (\xv-{\xv}_3)^2} e^{-s_1(\Vec{x}_1-\xv)^2} e^{-s_2(\Vec{x}_2-\xv)^2}
&=\pi^{d-1}\left(s_5(s_1+s_2)\right)^{\frac{1-d}{2}} e^{-\frac{s_1 s_2}{s_1+s_2}(\xv_1-\xv_2)^2}\,.
\end{align}}
Finally we do the $\xp, \xp_3$ integrals using  \eqref{expint} twice. This results in
\begin{align}
& \int_{0}^{\infty} \dd \xp \, \dd \xp_3 \ {\xp_3}^{2s-2\D_\f} e^{-s_5 (\xp-{x}_3^{\perp})^2 -s_1(x_1^{\perp}-\xp)^2-s_2(x_2^{\perp}-\xp)^2}\nn
&=\int_{-i \infty}^{i \infty} \frac{\dd \tau_1}{2\pi i} \frac{\dd \tau_2}{2\pi i} \,(-1)^{1-2\tau_2} \Gamma(\tau_1)\Gamma(\tau_2)\ s_5^{-\tau_1} \frac{\Gamma(2\tau_1+2\D_\f-2s-1)}{\Gamma(2\tau_1) \Gamma(2\tau_2)}\Gamma(2s-2\D_\f+1)\Gamma(2+2s-2\D_\f-2\tau_1)\nn
& \times (s_1 {\xp_1}+s_2{\xp_2})^{2(-\tau_1-\tau_2+s+1-\D_\f)} (s_1+s_2)^{2\tau_1+\tau_2+2(\D_\f-s-1)}e^{-\frac{s_1  s_2}{s_1+s_2}(x_1^{\perp}-x_2^{\perp})^2} 
{\Gamma(2(\tau_1+\tau_2+\D_\f-s-1))}\,.
\end{align}

We have now evaluated the bulk integrals. Next we evaluate the $s_1,s_2, s_5$ integrals. In order to evaluate those integrals we introduce two variables $\rho$ and $\lambda$ such that
\begin{align}
   1= \int_0^{\infty} \dd \lambda \ \d(\lambda-s_1-s_2-s_5)\int_0^{\infty} \dd \rho\, \d(\rho-s_1-s_2)\,,
\end{align}
and insert it in the integral. Then we rescale $s_i\rightarrow \rho \ {s}_i\,, i=1,2,5$ and $\lambda \rightarrow {\lambda} \rho$ and do the integrals over  $\rho, \lambda, {s}_5, {s}_2,s_1$ and $\tau_2$. 
This results in the following 
\begin{align}
\mathcal{I}^{(1)}_{tad} =&-\mathcal{P}_0\int  \frac{\dd s}{2\pi i} \frac{\dd s_4}{2\pi i} \frac{\dd\a}{2\pi i}\frac{\dd\tau_1}{2\pi i}\mathcal{B}(s,s_4,\a,\tau_1) \mathcal{F}_{s+\frac{1}{2}-\tau_1}(\xi).
\end{align}
where $\mathcal{F}_{s}(\xi), \mathcal{P}_0 ,\mathcal{B}(s,s_4,\a,\tau_1) $ are defined in \eqref{eqn:fxi} , \eqref{eqn:p0} and \eqref{eqn:b} respectively.

Finally, evaluating all the terms in \eqref{tad1} we obtain \eqref{eqn:resfinaltad}. %\PD{$\mathcal{B}=\mathcal{H}? $ in  \eqref{eqn:resfinal}}

\section{Details of bubble diagram}\label{bubble}

In this appendix, we give the details of evaluating the bubble diagram  \eqref{bub1}. This diagram contains contribution from total eight terms. We describe below how to evaluate one of those. The other diagrams can be evaluated in a similar fashion. Let us focus on the following term
\begin{align}
\mathcal{I}^{(1)}_{bub}=
\int dX\sqrt{g}\int dX_{3}\sqrt{g}\ G_{B\partial}(P_{1},X) \ G_{B\partial}(P_2, X_{3r}) \ G_{BB}(X_{3},X_{r})^2\,.
\end{align}
We use the Mellin-Barnes representation for the bulk-bulk propagator 
\begin{align}
G_{BB}^{2}(X_{3}, X_{r})& =\frac{(-1)^{2\Delta_{\phi}}}{(2\pi i)^2}\frac{\Gamma^{2}(c)}{\Gamma^{2}(a)\Gamma^{2}(b)}\int \dd s\int \dd s_{3}\frac{\Gamma(s)\Gamma(s_{3})\Gamma(a-s)\Gamma(a-s_{3})\Gamma(b-s)\Gamma(b-s_{3})}{\Gamma(c-s)\Gamma(c-s_{3})}\nn
& \times 4^{-s-s_{3}}\zeta^{s+s_{3}-2\Delta_{\phi}}
\end{align}
where 
\begin{align}
\zeta=\frac{\bar{\zeta}}{4yy_{3}}\,,\qquad 
\bar{\zeta}=(y-y_{3})^{2}+({\xv}-{\xv_{3}})^{2}+(x_{\perp}+x_{3\perp})^{2}\,.
\end{align}
We use the Schwinger parameterisation for 
\begin{align}
\bar{\zeta}^{s+s_{3}-2\Delta_{\phi}}=\frac{1}{\Gamma(2\Delta_{\phi}-s-s_{3})}\int_{0}^{\infty}\dd s_5\ s_{5}^{2\Delta_{\phi}-s-s_{3}-1}e^{-s_{5}((y-y_{3})^{2}+({\xv}-{\xv_{3}})^{2}+(x_{\perp}+x_{3\perp})^{2})} \,.
\end{align}
We  write the bulk-boundary propagators as described in \eqref{bulk_bdy1}. Let us first do the coordinate integrals
\begin{align}
\mathcal{I}_c &= \int dXdX_{3}\frac{(y y_3)^{\Delta_{\phi}}}{(yy_{3})^{s+s_{3}-2\Delta_{\phi}+d+1}} e^{-s_{1}(y^{2}+(\xv-\xv_{1})^{2}+(\xp-\xp_{1})^{2})}e^{-s_{2}(y_{3}^{2}+(\xv_{2}-\xv_{3})^{2}+(\xp_{2}+\xp_{3})^{2})}
\nn
&\times e^{-s_{5}((y-y_{3})^{2}+(\xv-\xv_{3})^{2}+(\xp+\xp_{3})^{2})}\nn
& =\mathcal{I}_y \mathcal{I}_v \mathcal{I}_x\,,
\end{align}
where $\mathcal{I}_y, \mathcal{I}_v ,\mathcal{I}_x$ refer to the $y, \xv, \xp$ integrals and are defined below.
The $y,y_{3}$ integrals result in 
\begin{align}
\mathcal{I}_y&=\int_{0}^{\infty}\dd y \dd y_{3}(yy_{3})^{3\Delta_{\phi}-s-s_{3}-d-1}e^{-s_{1}y^{2}-s_{2}y_{3}^{2}-s_{5}(y-y_{3})^{2}}\nn
&=\int \dd\tau \frac{4^{-\tau}(-1)^{d+s+s_3-3\D_\f-2\tau}\sqrt{\pi}}{ \G(\tau+1/2)} \G(3\D_\f-s-s_3-d) \G(d+s+s_3-3\D_\f+2\tau)
\nn & \times \G(3\D_\f-\tau-d-s-s_3)
\ s^{3\D_\f-2\tau-d-s-s_3}_5 (s_1 s_2+s_1 s_5+s_2 s_5)^{d+s+s_3-3\D_\f+\tau} \,.
\end{align}
The $\xv, \xv_3$ integrations result in
\begin{align}
\mathcal{I}_v&=\int \dd^{d-1}\xv \ \dd^{d-1}\xv_3 e^{-s_{1}(\xv-\xv_{1})^{2}}e^{-s_{2}(\xv_{2}-\xv_{3})^{2}}e^{-s_{5}(\xv-\xv_{3})^{2}}
=\frac{\pi^{d-1}}{(s_{2} s_1+s_2 s_5+{s_{1}s_{5}})^{\frac{d-1}{2}}}e^{-\frac{\st s_{2}}{\st+s_{2}}(\xv_{1}-\xv_{2})^{2}}\,,
\end{align}
where $\st=\frac{s_{1}s_{5}}{s_{1}+s_{5}}$.
Finally we are left with the  $\xp$and $\xp_3$ integrations 
\begin{align}
\mathcal{I}_x&=\int \dd\xp\int \dd\xp_3\ e^{-s_{1}(\xp_{1}-\xp)^{2}-s_{2}(\xp_{2}+\xp_{3})^{2}-s_{5}(\xp+\xp_{3})^{2}}\,.
\end{align}
The $\xp$ integral gives
\begin{align}
&\int_{0}^{\infty}\dd\xp \ e^{-(s_{1}+s_{5})\big(\xp+\frac{s_{5}\xp_{3}-s_{1} \xp_{1}}{s_{1}+s_{5}}\big)^{2}}e^{-\tilde{s}(\xp_{1}+\xp_{3})^{2}}
\nn
&=\int \dd\tau_{1}\Gamma(\tau_{1})(s_{1}+s_{5})^{-\tau_{1}}\big(\frac{s_{5}\xp_{3}-s_{1}\xp_{1}}{s_{1}+s_{5}}\big)^{1-2\tau_{1}}\frac{1}{2\tau_{1}-1}e^{-\tilde{s}(\xp_{1}+\xp_{3})^{2}} \,.
\end{align}
Now we use MB representation to do the $\xp_{3}$ integral, 
\begin{align}
(s_{5}\xp_{3}-s_{1}\xp_{1})^{1-2\tau_{1}}=\int \dd\sigma(s_{5}\xp_{3})^{\sigma}(-s_{1}\xp_{1})^{1-2\tau_{1}-\sigma}\frac{\Gamma(-\sigma)\Gamma(2\tau_{1}-1+\sigma)}{\Gamma(2\tau_{1}-1)} \,.
\end{align}
Now we do the $\xp_{3}$integration 
\begin{align}
&\int_{0}^{\infty}\dd\xp_{3}{\xp_3}^{\sigma}e^{-(\tilde{s}+s_{2})(\xp_3+\frac{\tilde{s} \xp_{1}+s2 \xp_{2}}{\tilde{s}+s_{2}})^{2}}e^{-\frac{\tilde{s}s_{2}}{\tilde{s}+s_{2}}(\xp_{1}-\xp_{2})^{2}}\nn
&
=\int \dd\tau_{2}\Gamma(\tau_{2})(\tilde{s}+s_{2})^{-1-\sigma+\tau_{2}}(s_{2}\xp_{2}+\tilde{s}\ \xp_{1})^{\sigma-2\tau_{2}+1}\frac{\Gamma(\sigma+1)\Gamma(-\sigma+2\tau_{2}-1)}{\Gamma(2\tau_{2})} e^{-\frac{\tilde{s}s_{2}}{\tilde{s}+s_{2}}(\xp_{1}-\xp_{2})^{2}} \,.
\end{align}
After integrating $\xp$and $\xp_{3}$ we are left with
\begin{align}
\mathcal{I}_x&=\int \dd\tau_{1}\int \dd\sigma\int \dd\tau_{2}\Gamma(\tau_{1})(s_{1}+s_{5})^{\tau_{1}-1}\frac{\Gamma(\sigma+1)\Gamma(-\sigma+2\tau_{2}-1)}{\Gamma(2\tau_{1})\Gamma(2\tau_{2})} \Gamma(-\sigma)\Gamma(2\tau_{1}-1+\sigma)\Gamma(\tau_{2})
\nn
&\times {s^{\sigma}_{5}}(-s_{1}\xp_1)^{1-2\tau_{1}-\sigma}(\tilde{s}+s_{2})^{-1-\sigma+\tau_{2}}(s_{2}\xp_{2}+\tilde{s} \xp_{1})^{\sigma-2\tau_{2}+1} e^{-\frac{\tilde{s}s_{2}}{\tilde{s}+s_{2}}(\xp_{1}-\xp_{2})^{2}} \,.
\end{align}
Now we perform $\sigma$ integration,
\begin{align}
    \mathcal{I}_x&=\int \dd\tau_{1}\int \dd\tau_{2}(-1)^{-2\tau_1}\Gamma(\tau_{1})(s_{1}+s_{5})^{\tau_{1}-1}\,(s_1\xp_1)^{2-2\tau_1}\,(s_2+\tilde{s})^{\tau_2}\, ({s_2 x_2^\perp+\tilde{s} x_1^\perp)}^{-2\tau_2}\,\nn & \times
     \frac{\G(\tau_2)}{s_5\,(2\tau_1-1)(-2+2\tau_1+2\tau_2)}\,\, _2F_1\left(1,2 {\tau_2};2 {\tau_1}+2 \tau_2-1;\frac{s_1 \tilde{s}\,x_1^\perp+s_1 s_2\,x_1^\perp}{s_2 s_5\, x_2^\perp+s_5 \tilde{s} \, x_1^\perp}+1\right)\,e^{-\frac{\tilde{s}s_{2}}{\tilde{s}+s_{2}}(\xp_{1}-\xp_{2})^{2}} \,.
\end{align}
So we have
\begin{align}
    \mathcal{I}^{(1)}_{bub}&=\mathcal{C}'\int \dd s\dd s_1 \dd s_2 \dd s_3 \dd s_5 \dd \tau_1 \dd \tau_2\, (-1)^{-2\tau_1}\mathcal{M}(s,s_1,s_2,s_3,s_5,\tau_1,\tau_2)e^{-\frac{\tilde{s}\,s_2}{\tilde{s}+s_2}(4 \xp_1\,\xp_2\,\xi)}
\end{align}
where
\begin{align}
 & \mathcal{M}(s,s_1,s_2,s_3,s_5,\tau_1,\tau_2) =\frac{4^{-s-s_{3}}\Gamma(s)\Gamma(s_{3})\Gamma(a-s)\Gamma(a-s_{3})\Gamma(b-s)\Gamma(b-s_{3})\G(\tau_1)\G(\tau_2)}{(\xp_1)^{2\tau_1-2}(2\tau_1-1)(-2+2\tau_1+2\tau_2)\Gamma(c-s)\Gamma(c-s_{3})\G(2\D_\f-s-s_3)}\mathcal{I}_y\nn
& \times \frac{s_1^{\D_\f+1-2\tau_1}\,s_2^{\D_\f-1}\,s_5^{2\D_\f-s-s_3-2}(s_{1}+s_{5})^{\tau_{1}-1}(s_2+\tilde{s})^{\tau_2}}{(s_1s_2+s_1s_5+s_2s_5)^{\frac{d-1}{2}}(s_2 x_2^\perp+\tilde{s} x_1^\perp)^{2\tau_2}}\,   _2F_1\left(1,2 {\tau_2},2 {\tau_1}+2 \tau_2-1;\frac{s_1 \tilde{s}\,x_1^\perp+s_1 s_2\,x_1^\perp}{s_5 s_2\, x_2^\perp+s_5 \tilde{s} \, x_1^\perp}+1\right) \,, \nn
  \mathcal{C}' &= \frac{(-1)^{2\Delta_{\phi}}\pi^{d-1}}{(2\pi i)^2}\frac{\Gamma^{2}(c)}{\Gamma^{4}(a)\Gamma^{2}(b)}\,.
\end{align}
Collecting the terms proportional to $e^{-\frac{\tilde{s}\,s_2}{\tilde{s}+s_2}(4 \xp_1\,\xp_2\,\xi)}$ in \eqref{bub1} we get the following 
\begin{align}\label{fin}
    \mathcal{I}^{\xi}_{bub}&=\mathcal{C}'\int \dd s\dd s_1 \dd s_2 \dd s_3 \dd s_5 \dd \tau_1 \dd \tau_2\,\bigg((-1)^{-2\tau_1}+(-1)^{1-2\tau_1-2\tau_2}+(-1)^{-2\tau_2}-1\bigg) \mathcal{M}(s,s_1,s_2,s_3,s_5,\tau_1,\tau_2)\nn & \times e^{-\frac{\tilde{s}\,s_2}{\tilde{s}+s_2}(4 \xp_1\,\xp_2\,\xi)}\,.
\end{align}

Now we evaluate the integral over $\tau_1$ and $\tau_2$ using the residue theorem. We have the following  poles from $\tau_1$
\begin{enumerate}[(i)]
    \item simple pole at $\tau_1=\frac{1}{2}$\,,
    \item simple pole at $\tau_1=1-\tau_2$ \,,
    \item  pole at $\tau_1=-n_1$ where $\quad n_1 \in \mathbb{Z}_{\geq 0}$ from  $\Gamma(\tau_1)$ \, .
\end{enumerate}
Subsequently, we collect the poles in $\tau_2$. After evaluating the corresponding residues, the expression in \eqref{fin} vanishes. The other terms in \eqref{bub1} can be evaluated using similar methods. It turns out that all the terms vanish after evaluating the integrals.

\bibliographystyle{JHEP}
\bibliography{HoloCFTs}
\end{document}